\documentclass[12pt,preprint]{aastex}
\shorttitle{}
\shortauthors{Nesvorn\'y et al.}
\begin{document}
\title{Young Solar System's Fifth Giant Planet?}
\author{David Nesvorn\'y}
\affil{Department of Space Studies, Southwest Research Institute,
1050 Walnut St., Suite 300, Boulder, Colorado 80302, USA} 

\begin{abstract}
Recent studies of solar system formation suggest that the solar system's giant planets 
formed and migrated in the protoplanetary disk to reach resonant orbits with all planets 
inside $\sim$15 AU from the Sun. After the gas disk's dispersal, Uranus and Neptune were 
likely scattered by gas giants, and approached their current orbits while dispersing the 
transplanetary disk of planetesimals, whose remains survived to this time in the region 
known as the Kuiper belt. Here we performed $N$-body integrations of the scattering phase 
between giant planets in an attempt to determine which initial states are plausible. 
We found that the dynamical simulations starting with a resonant system of four giant 
planets have a low success rate in matching the present orbits of giant planets, and 
various other constraints (e.g., survival of the terrestrial planets). The dynamical 
evolution is typically too violent, if Jupiter and Saturn start in the 3:2 resonance,
and leads to final systems with fewer than four planets. Several initial 
states stand out in that they show a relatively large likelihood of success in matching 
the constraints. Some of the statistically best results were obtained when assuming 
that the solar system initially had five giant planets and one ice giant, with the mass 
comparable to that of Uranus and Neptune, was ejected to interstellar space by Jupiter. 
This possibility appears to be conceivable in view of the recent discovery of a large 
number free-floating planets in interstellar space, which indicates that planet ejection 
should be common. 
\end{abstract}

\section{Introduction}
Studies of giant planets' interaction with a protoplanetary gas disk show that their orbits radially 
migrate, and typically achieve a compact configuration, in which pairs of the neighbor planets are 
locked in the mean motion resonances (Kley 2000, Masset \& Snellgrove 2001). The resonant planetary 
systems emerging from protoplanetary disks can become dynamically unstable after the gas disappears, 
leading to a phase when planets scatter off of each other. This model can explain the observed resonant 
exoplanets, commonly large exoplanet eccentricities (Weidenschilling \& Marzari 1996), and microlensing 
data that show evidence for a large number of planets that are free-floating in interstellar 
space (Sumi et al. 2011).

The solar system, with the widely spaced and nearly circular orbits of the giant planets, 
bears little resemblance to the bulk of known exoplanets. Yet, if our understanding of physics 
of planet--gas-disk interaction is correct, it seems likely that the young solar system 
followed the evolutionary path outlined above. Due to the convergent planetary migration in 
times before the gas disk's dispersal, each planet should have become trapped in a resonance 
with its neighbor. Jupiter and Saturn, for example, were most likely trapped in the 3:2 resonance 
(Masset \& Snellgrove 2001, Morbidelli \& Crida 2007, Pierens \& Nelson 2008), defined as 
$P_{\rm Saturn}/P_{\rm Jupiter}=1.5$, where $P_{\rm Jupiter}$ and $P_{\rm Saturn}$ are the orbital periods
of Jupiter and Saturn (this ratio is 2.49 today). 

To stretch to the present, more relaxed state, the outer solar system most likely underwent 
a violent phase when planets scattered off of each other and acquired eccentric orbits (Thommes 
et al. 1999, Tsiganis et al. 2005). The system was subsequently stabilized by damping the excess 
orbital energy into the transplanetary disk, whose remains survived to this time in the Kuiper belt. 
Finally, as evidenced by dynamical structures observed in the present Kuiper belt, planets 
radially migrated to their current orbits by scattering planetesimals (Malhotra et al. 1995, Levison 
et al. 2008).

\section{Method}

We conducted computer simulations of the early evolution of the solar system in an attempt 
to determine the initial states of planetary orbits (Batygin \& Brown 2010). First, we performed 
hydrodynamic and $N$-body simulations to identify the resonant configurations that may have occurred 
among the young solar system's giant planets. Our hydrodynamic simulations used Fargo (Masset 2000) 
and followed the method described in Morbidelli et al. (2007). As the Fargo simulations are 
CPU expensive, we used these results as a guide, and generated many additional resonant systems 
with the $N$-body integrator known as SyMBA (Duncan et al. 1998). 

Planets with masses corresponding to those of Jupiter, Saturn and ice giants were placed in initial
orbits with the period ratios slightly larger that those of the selected resonances. The planets 
were then migrated into resonances with the SyMBA code that was modified to include forces that 
mimic the effects of gas. We considered cases with four and five initial planets, where in the 
later case, an additional planet was placed into a resonant orbit between Saturn and the ice 
giants, or beyond the orbit of the outer ice giant. The fifth planet was given the mass between 
$3\times10^{25}$ and $3\times10^{26}$ g, which roughly corresponds to 1/3 and 3 times the mass of
Uranus (or Neptune). 

Different starting positions of planets, rates of the semimajor axis and eccentricity evolution, 
and timescales for the gas disk's dispersal produced different results. For Jupiter and Saturn,
we confined the scope of this study to the 3:2 and 2:1 resonances, because the former one is strongly 
preferred from previous hydrodynamic studies (Masset \& Snellgrove 2001, Morbidelli et al. 2007, Pierens 
\& Nelson 2008). The 2:1 resonance was included for comparison. The eccentricities and resonant amplitudes 
obtained here were typically $e<0.1$ and $<60^\circ$. The inner ice giant had $0.05<e<0.1$ in most cases, 
while the other planets acquired more circular orbits. 
 
The instability of a planetary system can occur after the gas disk's dispersal when the stabilizing effects 
of gas are removed. Such an instability can be triggered spontaneously (e.g., Weidenschilling \& Marzari 
1996), or by divergent migration of planets produced by their interaction with planetesimals leaking into 
the planet-crossing orbits from the transplanetary disk (Thommes et al. 1999). In the later case, the 
instability is produced when Jupiter and Saturn cross a major mean motion resonance (Tsiganis et al. 
2005; also known as the {\it Nice model}).

In the solar system, it is often assumed that the instability occurred at the time of 
the Late Heavy Bombardment (LHB) of the Moon some 3.9 Gy ago, when the lunar basins with
known ages formed (Hartmann et al. 2000). If so, the solar system's giant planets would be 
required to remain on their initial resonant orbits for about 700 Myr. To allow for this possibility, we 
sifted through the resonant configurations identified above and selected those that were stable
over $10^9$ yr, if considered in isolation. Only the stable systems were used for the follow-up 
simulations, in which we tracked the evolution of planetary orbits through and past the instability.

We included the effects of the transplanetary planetesimal disk in these simulations. The disk 
was represented by 1000 equal-mass bodies that were placed into orbits with low orbital eccentricities and 
inclinations, and in radial distances between $r_{\rm in}<r<r_{\rm out}$. The surface density was set to be 
$\Sigma=1/r$. The outer edge of the disk was placed at $r_{\rm out}=30$ AU, so that the planetesimal-driven 
migration is expected to park Neptune near its present semimajor axis. 

We considered cases with $r_{\rm in}=0.5$, 1, and 3.5 AU. The instability was triggered early for
$r_{\rm in}=0.5$ AU, as planetesimals leaked from the inner part of the disk into the planetary region, 
and interacted with planets. To trigger the instability for $r_{\rm in}=1$ and 3.5 AU, we broke 
the resonant locks by altering the mean anomaly of one of the ice giants. This method was inspired by the 
recent study of the late instability that showed that planets can slowly exchange orbital energy with 
a distant planetesimal disk, and break from the resonances when the resonant amplitude exceeds certain 
limits (Levison et al. 2011).

In either case discussed above, the scattering phase between planets starts shortly after the beginning 
of our simulations, which guarantees low CPU cost. We considered four different masses of the planetesimal disk, 
$m_{\rm disk}=10$, 20, 35, 50, 75 and 100 $M_{\rm Earth}$, and performed 30 simulations in each case, where 
different evolution histories were generated by randomly seeding the initial orbit distribution of planetesimals.
In total, we completed over 6000 scattering simulations. Each system was followed for 100 Myr, at which 
point the planetesimal disk was largely depleted and planetary migration ceased.

\section{Constraints}

We defined several criteria to measure the overall success of simulations. First 
of all, the final planetary system must have four giant planets (criterion A) with orbits that resemble 
the present ones (criterion B). Note that A means that one planet must be ejected in the 
five-planet runs, while all four planets need to survive in the four-planet runs. As for B, 
we claim success if the final mean semimajor axis of each planet is within 20\% to its present value, 
and if the final mean eccentricities and mean inclinations are no larger than 0.11 and 2$^\circ$, 
respectively. These thresholds were obtained by doubling the current mean eccentricity of Saturn 
($e_{\rm Saturn}=0.054$) and mean inclination of Uranus ($i_{\rm Uranus}=1.02^\circ$). 

For the successful runs, as defined above, we also checked on the history of encounters between giant planets, 
evolution of the secular $g_5$, $g_6$ and $s_6$ modes, and secular structure of the final planetary systems. 
To explain the observed populations of the irregular moons that are roughly similar at each planet (Jewitt 
\& Haghighipour 2007), all planets --including Jupiter-- must participate in encounters (Nesvorn\'y et al. 2007). 
Encounters of Jupiter and/or Saturn with ice giants may also be needed to excite the $g_5$ mode in the 
Jupiter's orbit to its current amplitude ($e_{55}=0.044$; Morbidelli et al. 2009). 

It turns out that it is generally easy to have encounters of one of the ice giants to Jupiter if planets
start in a compact resonant configuration. The amplitudes of $g_6$ and $s_6$ modes also do not pose a 
problem. It is much harder to excite $e_{55}$, however. We therefore opt for a criterion in which we require 
that $e_{55}>0.22$ in the final systems, i.e., at least half of its current value (criterion C). The 
$e_{55}$ amplitude was determined by following the final planetary systems for additional 10 Myr (without
planetesimals), and Fourier-analyzing the results (\v{S}idlichovsk\'y \& Nesvorn\'y 
1996).

The evolution of secular modes, mainly $g_5$, $g_6$ and $s_6$, is constrained from their effects
on the terrestrial planets and asteroid belt. As outer planets scatter and migrate, these frequencies 
change. This may become a problem, if $g_5$ slowly swipes over the $g_1$ or $g_2$ modes, because the strong 
$g_1=g_5$ and $g_2=g_5$ resonances can produce excessive excitation and instabilities in the terrestrial 
planet system (Brasser et al. 2009, Agnor \& Lin 2011). The behavior of the $g_6$ and $s_6$ modes, on the 
other hand, is important for the asteroid belt (Morbidelli et al. 2010). 

As $g_5$, $g_6$ and $s_6$ are mainly a function of the orbital separation between Jupiter and Saturn, the 
constraints from the terrestrial planets and asteroid belt can be conveniently defined in terms of 
$P_{\rm Saturn}/P_{\rm Jupiter}$. This ratio needs to evolve from $<$2.1 to $>$2.3 in $<1$ Myr 
(criterion D), which can be achieved, for example, if planetary encounters with an ice giant scatter Jupiter 
inward and Saturn outward. This condition may therefore require planetary encounters, but is more subtle 
in that not all simulations with Jupiter encounters are good. 

\section{Results}

\subsection{Four- and Five-Planet Results}

We start by discussing the results obtained with four planets that were assumed to be initially locked in the 
(3:2,3:2,4:3) resonances. The inner ice giant has the largest eccentricity ($e_3=0.06$). According 
to Levison et al. (2011), the system should therefore be driven toward the instability by the energy 
and momentum exchange between the inner ice giant and planetesimal disk. We set $r_{\rm in}=15$~AU, so 
that the inner disk edge is well beyond the orbits of the two ice giants ($a_3=9.6$AU and $a_4=11.6$ AU). 
This setup should be consistent with the late instability (Levison et al. 2011). 

The best results were obtained with $M_{\rm disk}=35M_{\rm Earth}$ and $M_{\rm disk}=50M_{\rm Earth}$. The 
fraction of simulations producing the final systems with four outer planets is 10\% and 13\%, respectively 
(criterion A). Only three of the total of 120 integrations performed for these disk masses satisfied 
our criterion B. This shows that it is very unlikely that the solar system evolved from these initial states.
The results do not improve when $M_{\rm disk}$ is varied. The light disks with $M_{\rm disk}\leq20M_{\rm Earth}$ 
lead to final systems with fewer than 4 planets. The heavy disks with $M_{\rm disk}>50 M_{\rm Earth}$ do not 
work as well, as we discuss in more detail below (\S4.2).

We compare these results with those obtained for the five-planet systems. To start with, the five planets 
were placed into the (3:2,3:2,4:3,5:4) resonances and fifth planet was given the mass equal to that of Uranus. 
As in the four-planet case, we fix $r_{\rm in}=15$~AU so that the most eccentric inner ice giant ($e_3=0.07$) 
has about the same radial distance from the inner edge of the disk.

\begin{figure}
\epsscale{0.6}
\plotone{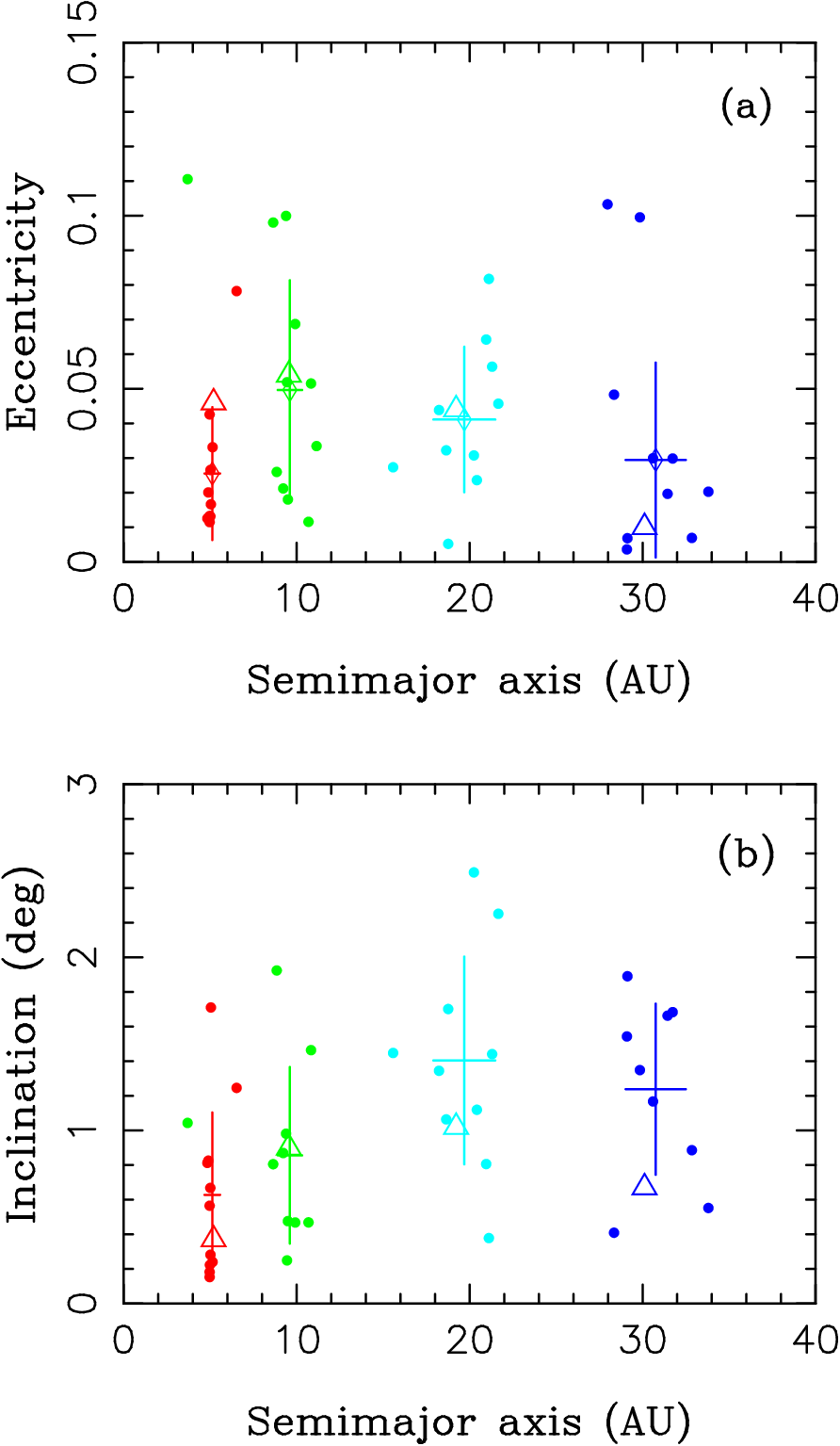}
\caption{Final orbits obtained with five planets starting in the (3:2,3:2,4:3,5:4) resonances,
$M_{\rm disk}=50M_{\rm Earth}$, and $r_{\rm in}=15$ AU. Only the systems ending with four planets are 
plotted (dots). The error bars show the average and 
rms of the orbital elements. The mean orbits of Jupiter, Saturn, Uranus and Neptune are shown by red, 
green, turquoise and blue triangles.}
\label{orb}
\end{figure}

\begin{figure}
\epsscale{0.6}
\plotone{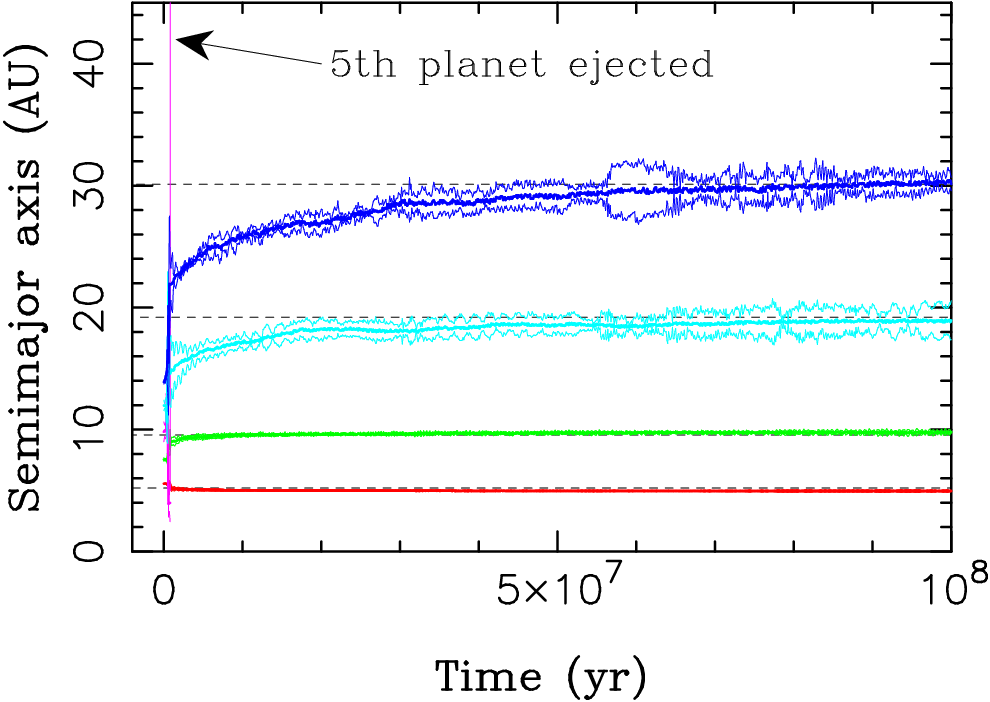}
\caption{The orbit histories of giant planets in one of the simulation with five initial planets. 
In this case, the five planets were started in the (3:2,3:2,4:3,5:4) resonances, 
$M_{\rm disk}=50M_{\rm Earth}$, and $r_{\rm in}=15$ AU. After series of encounters with Jupiter, 
the inner ice giant was ejected from the solar system at $8.2\times10^5$ yr (purple path). The 
remaining planets were stabilized by the planetesimal disk, and migrated to orbits that very 
closely match those of the outer planets (dashed lines).}
\label{mig}
\end{figure}

The best five-planet results were obtained for $M_{\rm disk}=50M_{\rm Earth}$ with the fraction of systems matching 
criteria A and B raising to 37\% and 23\%, respectively (Fig. \ref{orb}). This shows that it 
is it roughly ten times more likely to obtain a good solar system analog starting from five planets than from 
four planets, at least in the case discussed above. This result is not unexpected because systems starting from a 
compact resonant configuration typically suffer a rather violent instability, and tend to loose planets
(Fig. \ref{mig}).

\subsection{Effects of $M_{\rm disk}$}

The results for the four-planet case do not improve when the mass of the planetesimal disk is increased. 
With $M_{\rm disk}=100M_{\rm Earth}$, Neptune migrates too far and/or the divergent migration of Jupiter and 
Saturn moves these planets too far apart so that $2.8<P_{\rm Saturn}/P_{\rm Jupiter}<3.2$ in the end. 
The former problem can be resolved if the disk were truncated at $\simeq$25 AU, but we do not see any 
obvious cure to the second problem. This is because the final radial spacing of the Jupiter's and Saturn's 
orbits depends on the mass of material processed through $<$10 AU, which is simply too large for 
$M_{\rm disk}=100M_{\rm Earth}$. In addition, Saturn and Jupiter tend to end up on unrealistically circular 
orbits for large $M_{\rm disk}$, because their eccentricities are damped by the large mass that these 
planets interact with.

\subsection{Effects of $r_{\rm in}$}

The success rate of simulations depends on $r_{\rm in}$ with the larger $r_{\rm in}$ values 
leading to a lower success rate. For example, the best four-planet case discussed above matched 
the criterion B in 10\% of cases for $r_{\rm in}=12.6$ AU and $M_{\rm disk}=50M_{\rm Earth}$, which is 
only slightly lower that the success rate obtained for the five-planet case with $r_{\rm in}=17.5$ AU 
and $M_{\rm disk}=50M_{\rm Earth}$. On the other hand, the success rate of the four-planet case drops 
to $\lesssim$3\% if $r_{\rm in}=17.5$ AU.

The sensitivity of the success rate to $r_{\rm in}$ stems from the following. With the small $r_{\rm in}$ 
values, planetesimals can rapidly leak from the inner part of the transplanetary disk into the planetary 
region, and stabilize planets very soon after the onset of the instability. With large $r_{\rm in}$, on the 
other hand, the system lacks the stabilizing effect of planetesimals on planet-crossing orbits, planetary
orbits become increasingly excited, and planets can be ejected. This highlights the importance of 
planet-crossing population of planetesimals on the results.

\subsection{Mass of the Fifth Planet}

We tested cases with different masses of the fifth planet. Following the mass gradient, we placed 
the fifth planet with the mass intermediate between those of Saturn and Uranus into an exterior 
resonance with Saturn, or a planet with mass lower than that of Neptune beyond the orbit of Neptune. 
The former case did not lead to any improvement of the results discussed above, because the system 
became to violent and frequently ejected the less massive planets from the system. In the later case, 
the statistics also did not improve because the inner ice giant got ejected, thus leaving a final 
system with incorrect masses. 

\subsection{Initial Resonant Configuration}

We found that the behavior of planetary systems, and whether or not the final systems end up resembling 
the outer solar system, do not depend in detail on the initial resonant sequence between Saturn and 
ice giants. Instead, the results are mainly sensitive to the initial resonance between Jupiter 
and Saturn, with 3:2 or 2:1 cases considered here, and the overall initial spread of the ice giants' 
orbits. 

As for the 3:2 Jupiter-Saturn resonance, the case with (3:2,3:2,4:3,5:4) discussed above was one of the 
most compact resonant systems studied here. Other compact systems show similar success rates. 
The most relaxed system studied here was (3:2,3:2,3:2,3:2)$=$(3:2)$^3$. With $r_{\rm in}=17.5$ AU
($a_5=16.1$ AU in this case) and $M_{\rm disk}=35M_{\rm Earth}$, the success rate was higher than the one 
obtained for a similar setup with (3:2,3:2,4:3,5:4). For example, the criterion A was satisfied in 33\% 
of simulations with (3:2)$^3$, while it was 13\% in the former case, and criterion B was satisfied in 
17\% of simulations. This shows that a more relaxed initial resonant configuration of ice giants 
can improve the statistics (mainly for the lower mass disks; $M_{\rm disk}\lesssim 35M_{\rm Earth}$).  

The four-planet case with the 2:1 Jupiter-Saturn resonance shows a relatively large success rate in 
matching our criteria A and B. The instability produced in these simulations tends to be weaker than
in the 3:2 case, which seems to have positive impact on the results. There are fewer cases of
Jupiter encounters with one of the ice giants, however, leading to orbit histories that 
violate constraints C and D. We did not found any case, for (2:1,3:2,3:2) and $M_{\rm disk} =
35M_{\rm Earth}$, where these criteria would be satisfied. 

The five-planet case with the 2:1 Jupiter-Saturn resonance is interesting. Better results
were obtained in this case when the inner ice giant had lower mass, because when Jupiter and Saturn
start in the 2:1 resonance, their period ratio needs to change by $\sim$0.5 only (from 2 to 2.49), which
requires a smaller perturbation. Assuming a planet half the Uranus mass, $20 \leq M_{\rm disk} \leq 
35 M_{\rm Earth}$ and $r_{\rm in}=a_5+1$ AU, where $a_5$ denotes the semimajor axis of the outer 
ice giant, criterion A was satisfied in $\sim$50\% of cases and criterion B was satisfied in 
20-30\% of cases. These results are pretty much independent of the initial resonant sequence between 
Saturn and the ice giants. All successful jobs show Jupiter encounters and have $\sim10$\% chance of 
simultaneously matching criteria C and D as well (\S4.8).

\subsection{Planetary encounters}

Only $\sim$3\% of simulations performed for the four-planet case with (3:2,3:2,4:3), $r_{\rm in}=12.6$ AU, 
and $M_{\rm disk} = 50 M_{\rm Earth}$, satisfy  criteria A and B, {\it and} show encounters of one of the 
ice giants to Jupiter. The statistics for other resonances is similarly low. Consequently, in absence 
of Jupiter encounters, Jupiter and Saturn end up too close to each other if $M_{\rm disk} < 50 M_{\rm Earth}$, 
and their orbits are too circular.

\begin{figure}
\epsscale{0.6}
\plotone{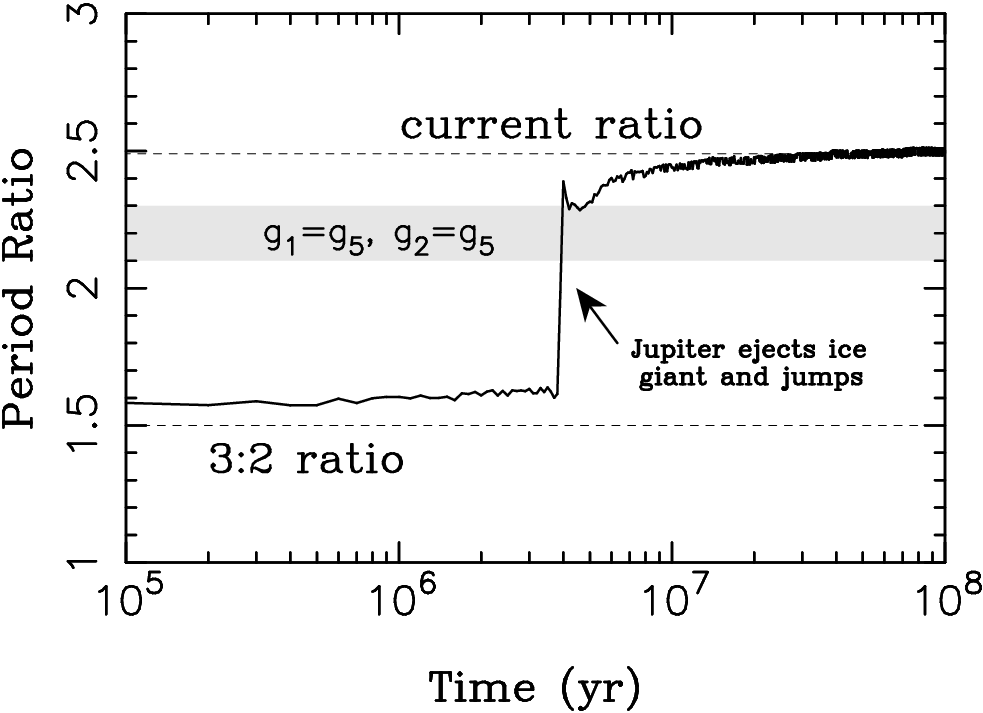}
\caption{Period ratio, $P_{\rm Saturn}/P_{\rm Jupiter}$, for a selected five-planet case with (3:2,3:2,4:3,5:4)
and $M_{\rm disk}=35M_{\rm Earth}$. The fifth planet was ejected by Jupiter at 3.5 Myr. This produced
a jump of $P_{\rm Saturn}/P_{\rm Jupiter}$ from $\sim 1.5$ to 2.4. The kind of evolution shown here, known as 
the jumping Jupiter (Morbidelli et al. 2010), may be needed to avoid the region at 2.1-2.3, where the
$g_1=g_5$ and $g_2=g_5$ resonances occur (Brasser et al. 2009, Agnor \& Lin 2011).}
\label{per}
\end{figure}

Larger disk masses produce more plausible final period ratios, as Jupiter and Saturn can slowly separate
from each other by scattering planetesimals, but these evolution paths do not satisfy criterion D. 
Overall, the criterion D was satisfied only in $\sim$1\% of the four-planet cases, which is troubling. 
For a comparison, practically in all five-planet simulations reported here, all planets, 
including Jupiter, participate in planetary encounters. For example, the criterion D was satisfied in 
50\% of cases that also satisfied A and B for (3:2,3:2,4:3,5:4), $M_{\rm disk} = 50 M_{\rm Earth}$ and 
$r_{\rm in}=15$ AU (Fig. \ref{per}), and in over 60\% of good cases for (2:1,3:2,3:2,4:3), $M_{\rm disk} = 
20 M_{\rm Earth}$ and $r_{\rm in}=19$ AU. 

\subsection{Secular Modes}

We analyzed our simulations to determine the fraction of cases in which the secular structure of 
the final systems was similar to that of the present solar system. We did not use the Jupiter's encounters 
as a proxy for good cases (Batygin \& Brown 2010), because we found that most simulations with 
Jupiter's encounters did not satisfy our criterion C ($e_{55}>0.22$). In fact, 
the overall success rate for the criterion C was disappointingly low, both for 
the four- and five-planet systems. In most cases, the initial eccentricity of Jupiter, or the one excited by 
planetary encounters, was damped by planetesimals passing through $<10$ AU, so that $e_{55}\lesssim0.01$   
in the end.

This problem could most easily be resolved for low initial masses of the planetesimal disk, because in this 
case, the $e_{55}$ amplitude --initial or produced by the planetary encounters-- will most likely survive. 
As planets can easily be ejected for low $M_{\rm disk}$, more than four initial planets will probably be 
required. Still, to make things work for the promising five-planet case, the success in matching the 
criteria A and B should be improved for low $M_{\rm disk}$, because the systems studied here with 
$M_{\rm disk}=20 M_{\rm Earth}$ were generally too violent. This could be achieved, for example, by including 
some sort of dissipation in the planetesimal disk, possibly produced by the collisions between planetesimals.

\section{Conclusions}

The formation of Uranus and Neptune is a long-standing problem in planetary science, because their
accretion at $\sim$20 and $\sim$30 AU would require implausibly long timescales (Safronov 1969, Levison
\& Stewart 2001). The ice giants can form more easily at $<$15 AU, where densities are higher and 
dynamical timescales are shorter (e.g., Robinson \& Bodenheimer 2010, Jakub\'{\i}k et al. 2011).
Our five-planet resonant systems start with all ice giant at $<$15 AU, if Jupiter and Saturn are
in the 3:2 resonance. If Jupiter and Saturn start in the 2:1 resonance, on the other hand, the whole 
planetary system is more spread, and the outer ice giant is at $\sim$18-20 AU, where its formation
can be problematic. Future work will need to address these issues.

\acknowledgments
This work was supported by NLSI and NSF.

\clearpage

\clearpage

\clearpage

\end{document}